%% file: brau_dpf2015.tex
\documentclass[12pt]{article}
\usepackage{graphicx}
\usepackage{hyperref}

\newcommand\pubnumber{DPF2015-195}
\newcommand\pubdate{\today}

\def\oregon{Center for High Energy Physics\\
University of Oregon, Eugene 97403-1274}
\def\support{\footnote{Work supported by the U.S. Department of Energy,
Office of Science, Office of High Energy Physics.}}

\def\Title#1{\begin{center} {\Large #1 } \end{center}}
\def\Author#1{\begin{center}{ \sc #1} \end{center}}
\def\Address#1{\begin{center}{ \it #1} \end{center}}

\newcommand\pubblock{\rightline{\begin{tabular}{l} \pubnumber\\
         \pubdate  \end{tabular}}}
\newenvironment{Abstract}{\begin{quotation}  }{\end{quotation}}
\newenvironment{Presented}{\begin{quotation} \begin{center} 
             PRESENTED AT\end{center}\bigskip 
      \begin{center}\begin{large}}{\end{large}\end{center} \end{quotation}}
\def\Acknowledgments{\bigskip  \bigskip \begin{center} \begin{large}
             \bf ACKNOWLEDGMENTS \end{large}\end{center}}
\input econfmacros.tex
\begin{document}
\begin{titlepage}
\pubblock

\vfill
\Title{500 GeV ILC Operating Scenarios}
\vfill
\Author{ James E. Brau\support}
\Address{\oregon}
\begin{center} {representing}
\end{center}
\Author{ILC Parameters Joint Working Group}
\Author{T. Barklow (SLAC), J. Brau (Oregon), K. Fujii (KEK), J. Gao (IHEP), J. List (DESY), N.Walker (DESY), K. Yokoya (KEK)}
\vfill
\begin{Abstract}
The ILC Technical Design Report documents the design of a 500
GeV linear collider, but does not specify the center-of-mass energy steps of operation
for the collider. The ILC Parameters Joint Working Group has studied possible running
scenarios, including a realistic estimate of the real time accumulation of integrated luminosity
based on ramp-up and upgrade processes, and considered the evolution of the physics outcomes.
These physics goals include Higgs precision measurements, top quark measurements
and searches for new physics. We present an ``optimized" operating scenario and the anticipated
evolution of the precision of the ILC measurements.
\end{Abstract}
\vfill
\begin{Presented}
DPF 2015\\
The Meeting of the American Physical Society\\
Division of Particles and Fields\\
Ann Arbor, Michigan, August 4--8, 2015\\
\end{Presented}
\vfill
\end{titlepage}
\def\thefootnote{\fnsymbol{footnote}}
\setcounter{footnote}{0}

\section{Introduction}
The ILC Technical Design Report (TDR)~\cite{ilc-tdr} provides a blueprint
for the ILC program, based on
many years of a globally coordinated R\&D program.
This realistic technical design and implementation plan
has been optimized for performance, cost and risk.
The R\&D program included:
\begin{itemize}
\item construction and commissioning of superconducting RF test
facilities for accelerators all over the world,
\item improvement in accelerating cavities production processes, and
\item plans for mass production of 16,000 superconducting
cavities needed to drive the ILC's particle beams.
\end{itemize}
The TDR includes details for two state-of-the-art detectors
(SiD and ILD)
operating in a push-pull configuration, as well as an extensive
outline of the geological and civil engineering studies
conducted for siting the ILC.

\begin{figure}[htb]
\centering
\includegraphics[height=4in,trim={0 1.5cm 0 0},clip]{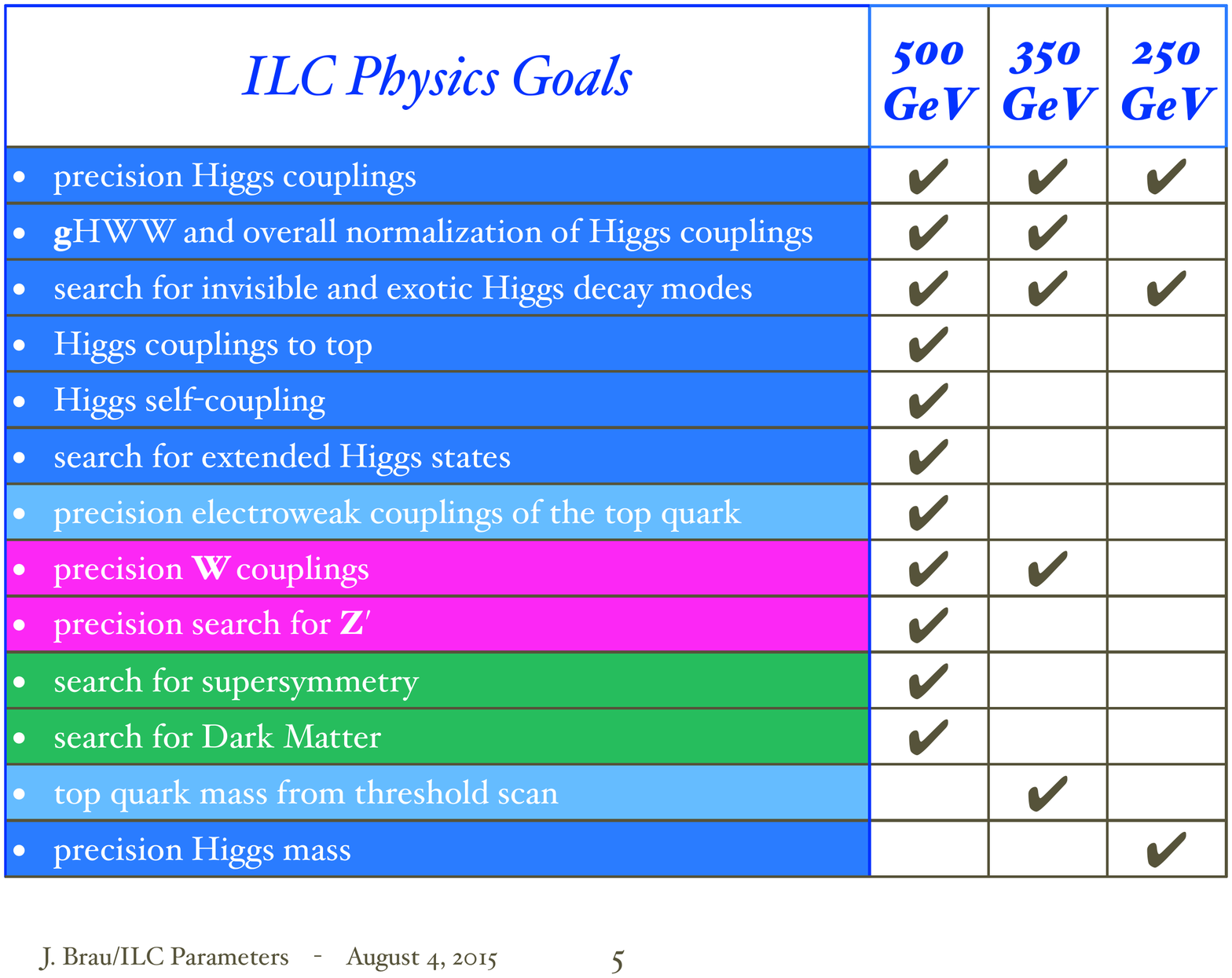}
\caption{ILC Physics Goals.}
\label{fig:physics}
\end{figure}

The physics program envisioned for the 500 GeV ILC
is rich (Figure~\ref{fig:physics}), with the collider 
operating at different center-of-mass energy points
to optimize physics outcomes.
Operations will start at the full
center-of-mass energy of $500$\,GeV, followed by
$250$ and $350$\,GeV running, for an initial total of eight to ten years.
The collider luminosity will then be upgraded  
for intense running for about another ten years.

\section{Running Scenarios}

While the TDR specifies the upper energy of 500 GeV for the
initial phase of the ILC, there is flexibility in choosing the
operating energy; this is one of the strengths of the ILC.
In order to plan for optimized operations, considering machine and
physics issues, a working group was charged by the Linear Collider Collaboration (LCC)~\cite{lcc}
to compare various running scenarios for a 500 GeV ILC and to recommend a standard
set of total integrated luminosities for use in physics studies.  It is
recognized that the actual running scenario will depend on many
future factors, including physics results of the LHC and the ILC.

The basis for the scenarios considered was the 
TDR baseline, emphasizing the upper energy reach with maximum
discovery potential, and assuming 20 years of operation.
Many scenarios were compared and contrasted.  Here we present
the three which were used in the final phase of the study,
as well as a fourth based on the Snowmass white paper~\cite{snowmass},
although
the Snowmass scenario assumes 15 years of operation.
This study is presented elsewhere in more detail~\cite{scenarios}.

The detailed assumptions for this study were:
\begin{itemize}
\item A full calendar year is assumed to be 8 months at a 75\% efficiency (the RDR~\cite{rdr}
assumption). This corresponds to Y = $1.6 \times 10^7$ seconds of integrated
running, significantly higher than a Snowmass year of $10^7$ seconds.
\item A ramp-up of luminosity performance is assumed where expected.
%\begin{itemize}
%\item (a) initial construction and after ‘year 0’ commissioning;
%\item (b) after a downtime for a luminosity upgrade;
%\item (c) a change in operational mode which may require some learning
%curve (e.g. going to 10-Hz collisions).
%\end{itemize}
\begin{itemize}
\item For the initial physics run after construction and year 0 commissioning, the RDR
ramp of 10\%, 30\%, 60\% and 100\% is assumed over the first four years.
\item The ramp after the shutdowns for installation of the luminosity upgrade is
assumed to be slightly shorter (10\%, 50\%, 100\%) with no year 0.
\item Going down in center-of-mass energy from 500 GeV to 350 GeV or 250 GeV is
assumed to have no ramp, since there is no machine modification.
\item Going to 10-Hz operation at 50\% gradient does assume a ramp (25\%, 75\%,
100\%), since 10-Hz affects the entire machine.
\end{itemize}
\item A major 18-month shutdown is assumed for the luminosity upgrade.
\item Unlike TDR, 10-Hz and 7-Hz operation is assumed at 250 GeV and 350 GeV.
\end{itemize}

\noindent The physics reach of the ILC program depends on the
total integrated luminosities collected at various center-of-mass energies,
as well as the various beam polarization combinations collected
at each of those energies.
The highest achievable degree of polarization is desirable, 
and the TDR presents the assumed polarizations
of P(e$^-$) = 80\% and P(e$^+$) = 30\%
(higher values are possible for both species). 
The choice of combinations results from the dependence of 
processes on the polarization:
\begin{itemize}
\item s-channel Z/$\gamma$ is allowed for e$^-_L$ e$^+_R$ \& e$^-_R$ e$^+_L$, where Z favors e$^-_L$ e$^+_R$,
\item t-channel Z/$\gamma$ is allowed for e$^-_L$ e$^+_L$ and e$^-_R$ e$^+_R$,
\item BSM t-channel allows like sign helicities, but W or $\nu_e$ t-channel
exchange is allowed only for unlike sign helicities.
\end{itemize}
Table~\ref{tab:pols} presents the fractions of integrated luminosity
collected for each center-of-mass energy based on these considerations.

\begin{table}[t]
\begin{center}
\begin{tabular}{|l|c|c|c|c|}
\hline
  &fraction with (P(e$^-$),P(e$^+$))= &&&  \\
 &  (L,R) &  (R,L) &  (L,L)  & (R,R) \\ \hline
 $\sqrt{s}$ & [\%] & [\%] & [\%] & [\%] \\ \hline
 250 GeV  &  67.5  &    22.5   &   5  &  5 \\
 350 GeV  &  67.5  &    22.5   &   5  &  5 \\ 
 500 GeV  &  40    &  40  &  10  &  10 \\ \hline
\end{tabular}
\caption{Relative sharing between beam helicity configuration versus 
center-of-mass energy.}
\label{tab:pols}
\end{center}
\end{table}

\begin{figure}[htb]
\centering
\includegraphics[height=2in]{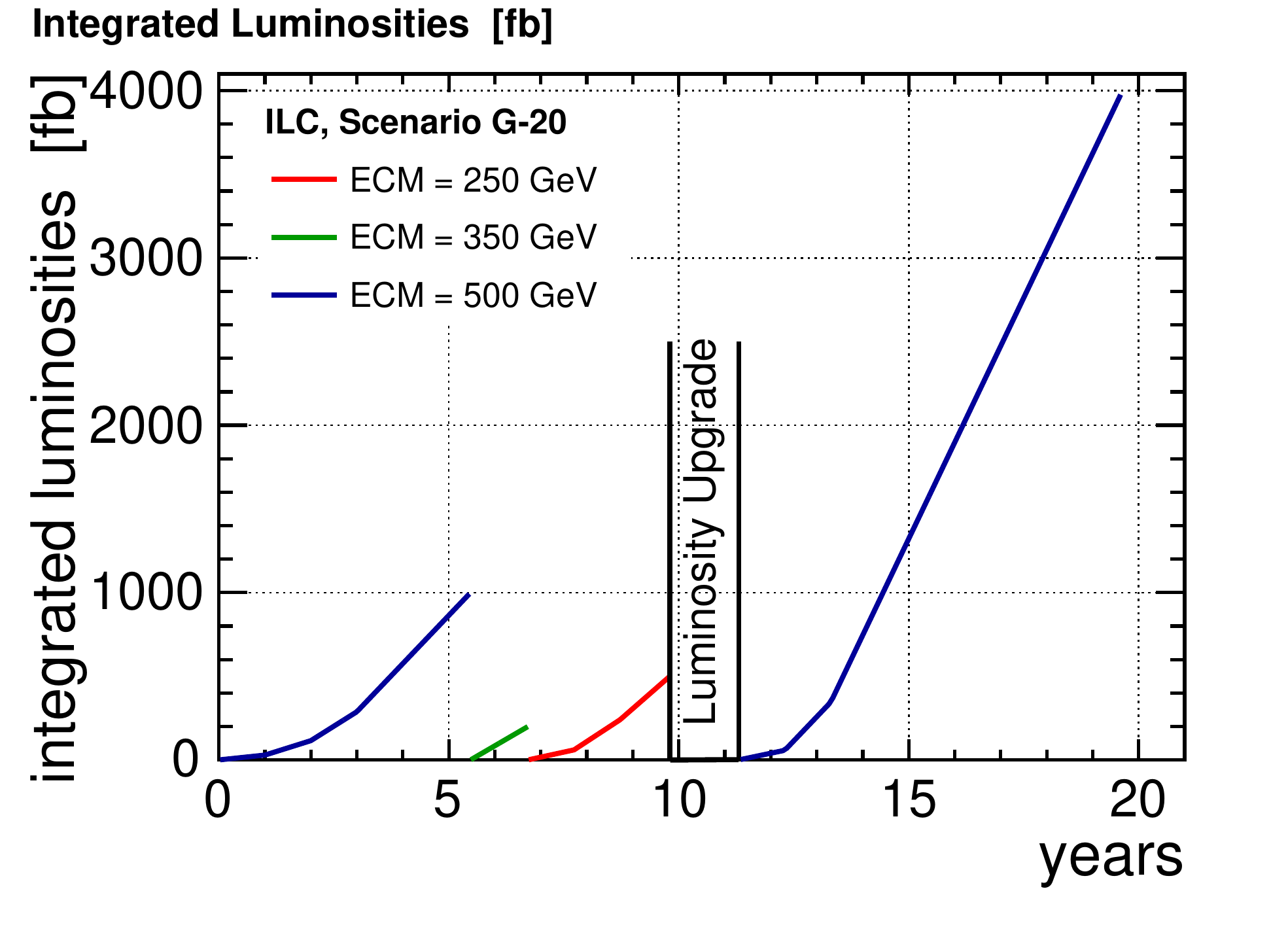}
\includegraphics[height=2in]{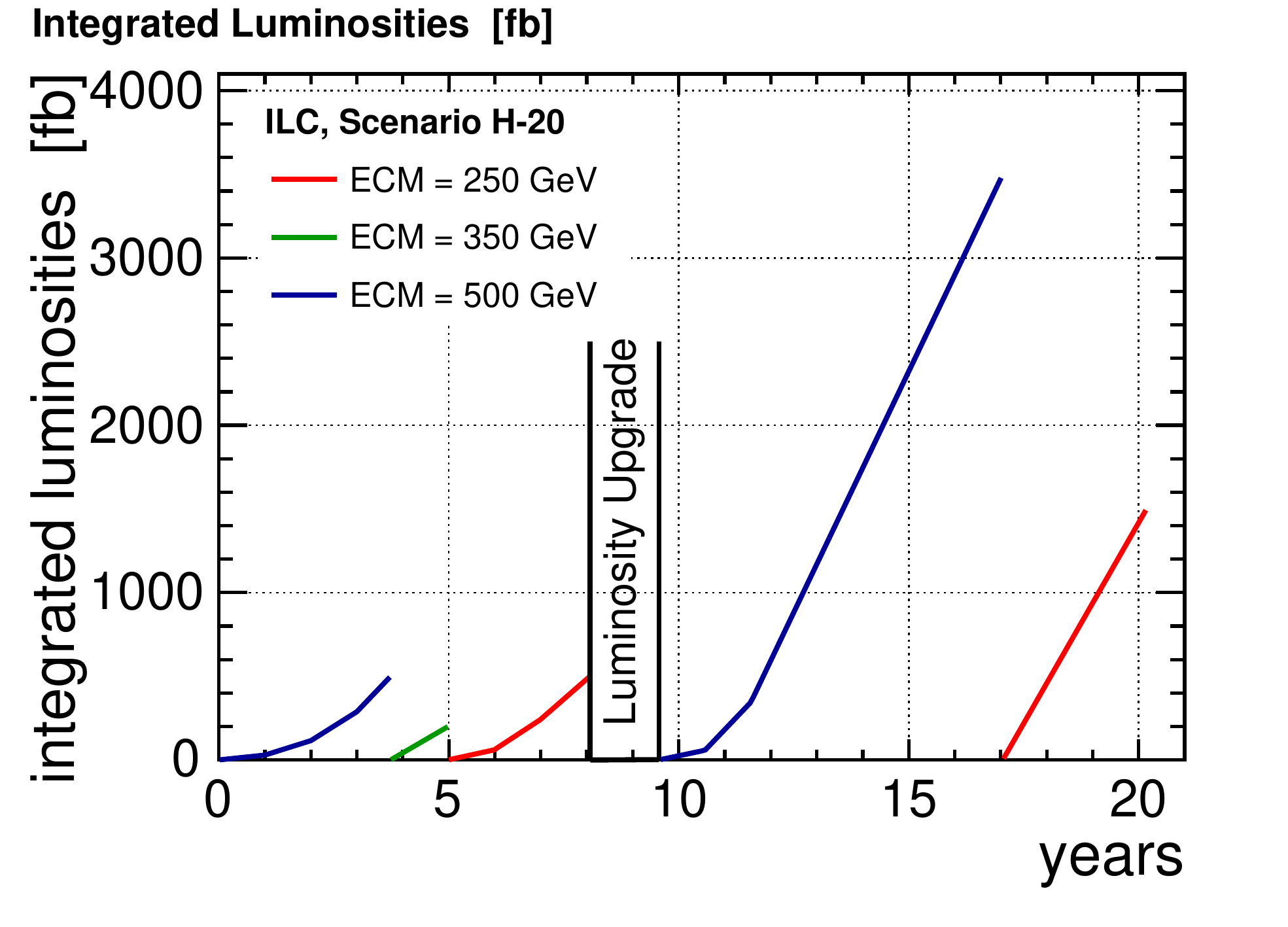}
\includegraphics[height=2in]{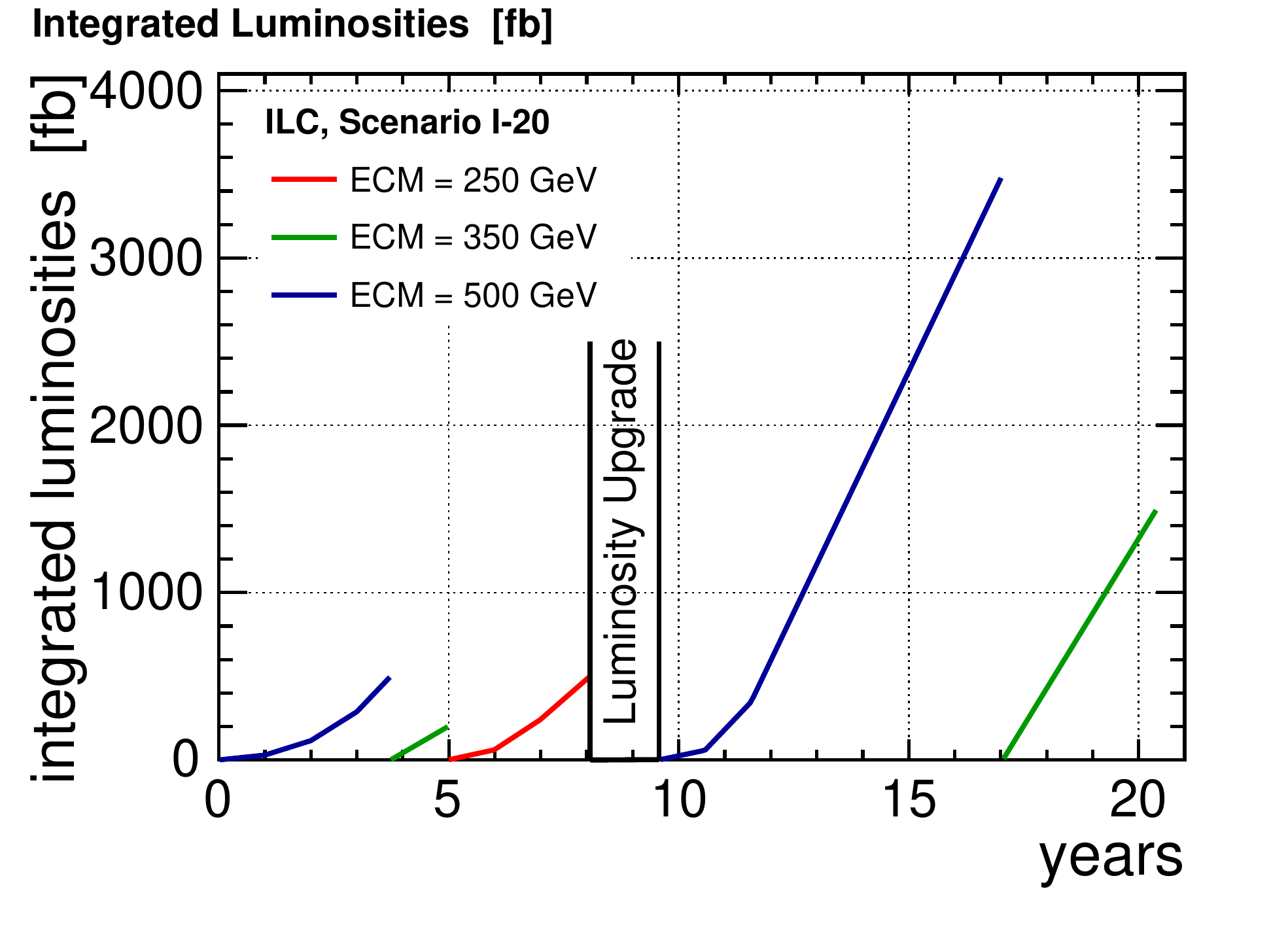}
\includegraphics[height=2in]{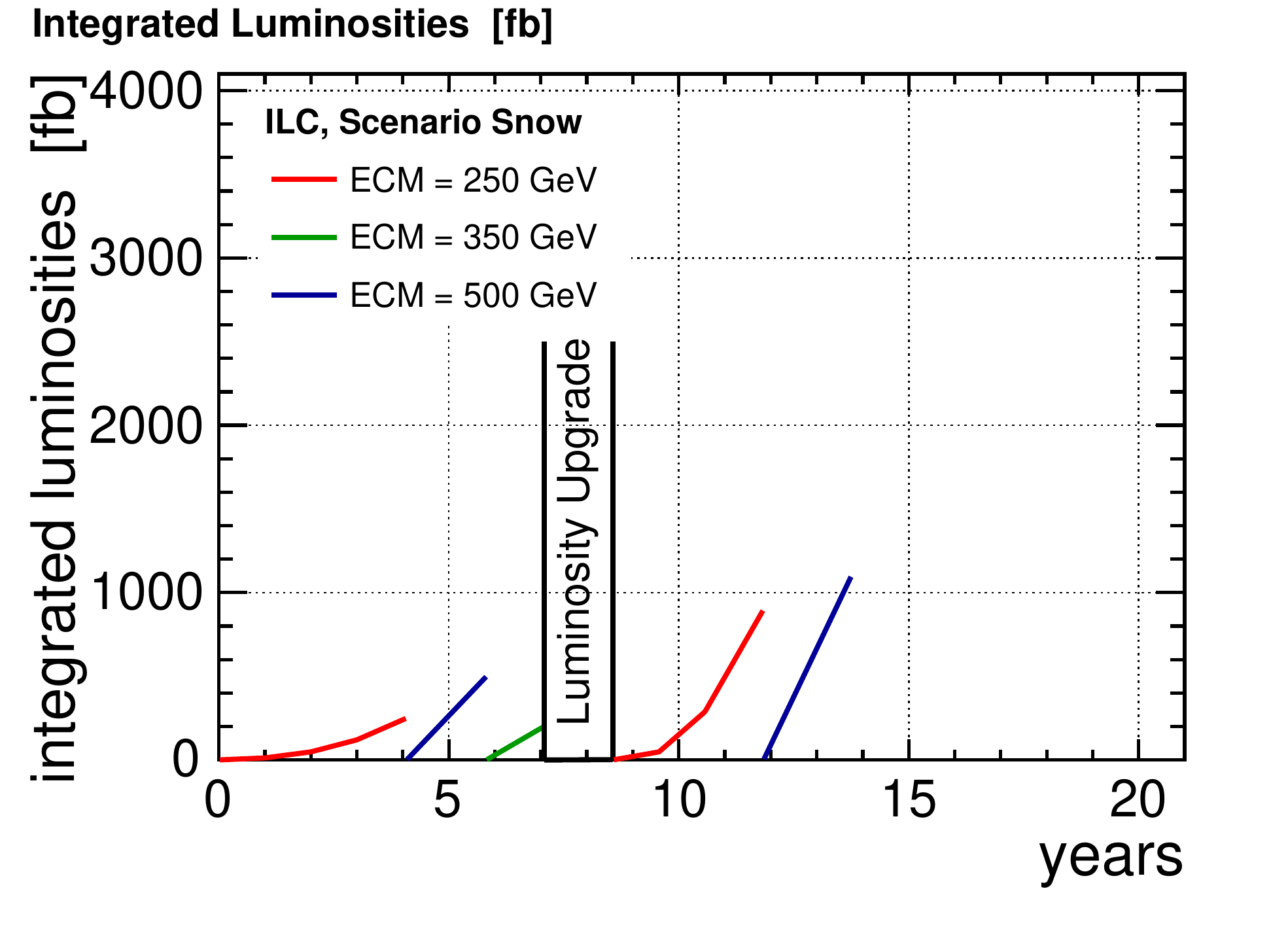}
\caption{Integrated luminosities for  the G-20, H-20, I-20, and Snowmass
scenarios.}
\label{fig:lums}
\end{figure}

Figure~\ref{fig:lums} presents the assumed progression of integrated
luminosities for the three scenarios (G-20, H-20, and I-20) and the
Snowmass case (Snow). In each case, a luminosity upgrade is planned
after eight to ten years.

\section{Higgs boson}

\begin{figure}[htb]
\centering
\includegraphics[height=2.1in]{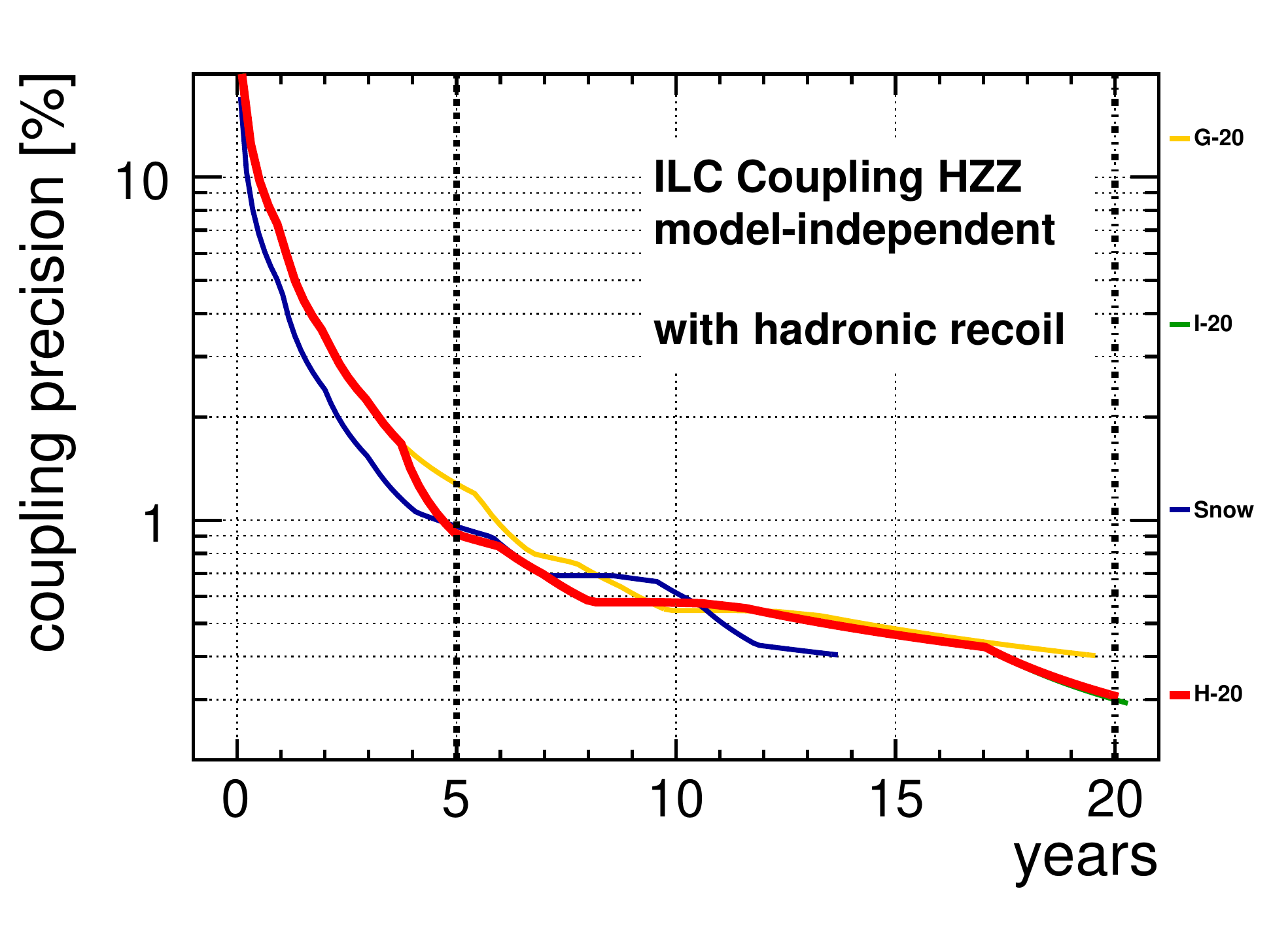}
\includegraphics[height=2.1in]{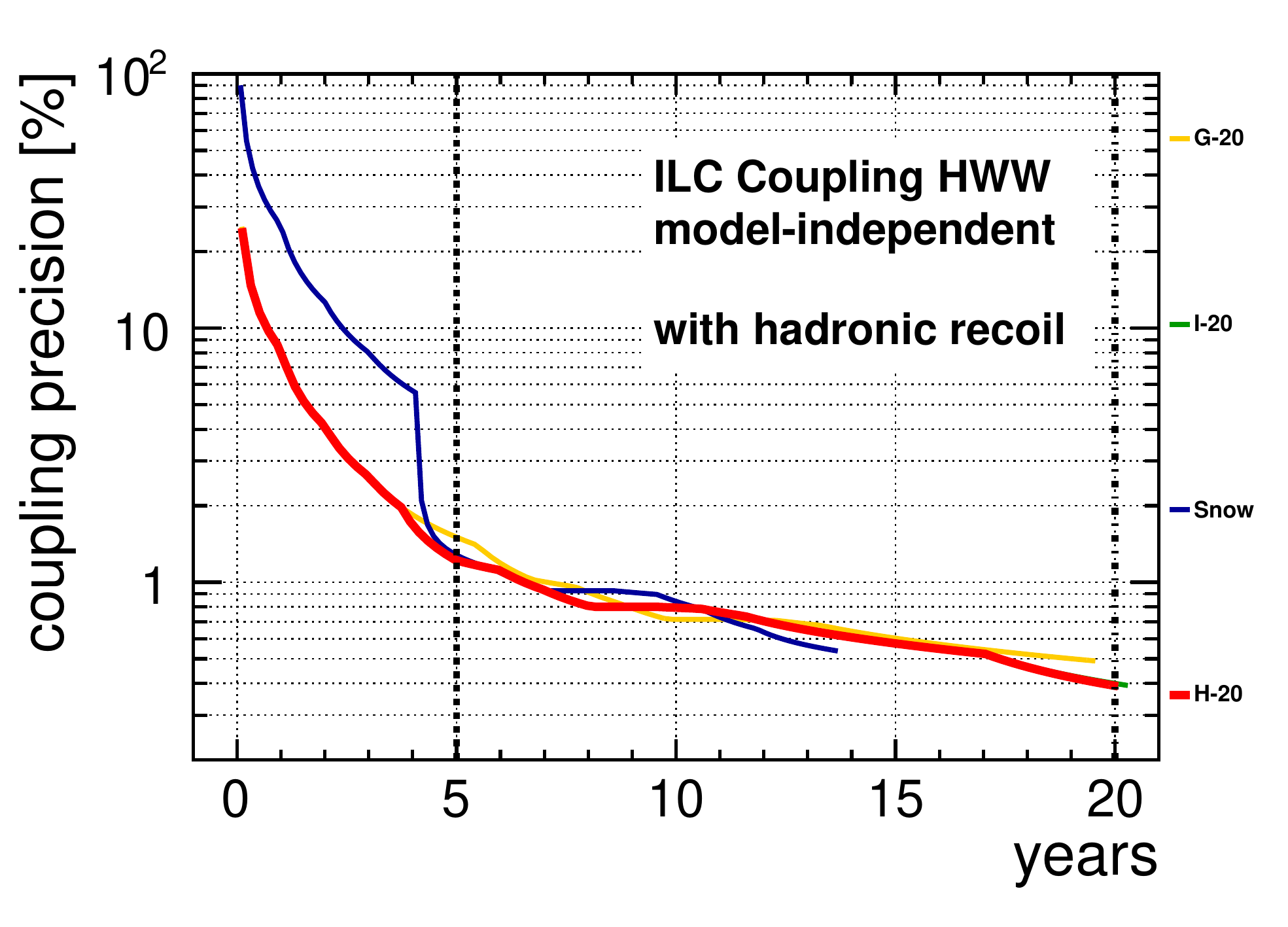}
\includegraphics[height=2.1in]{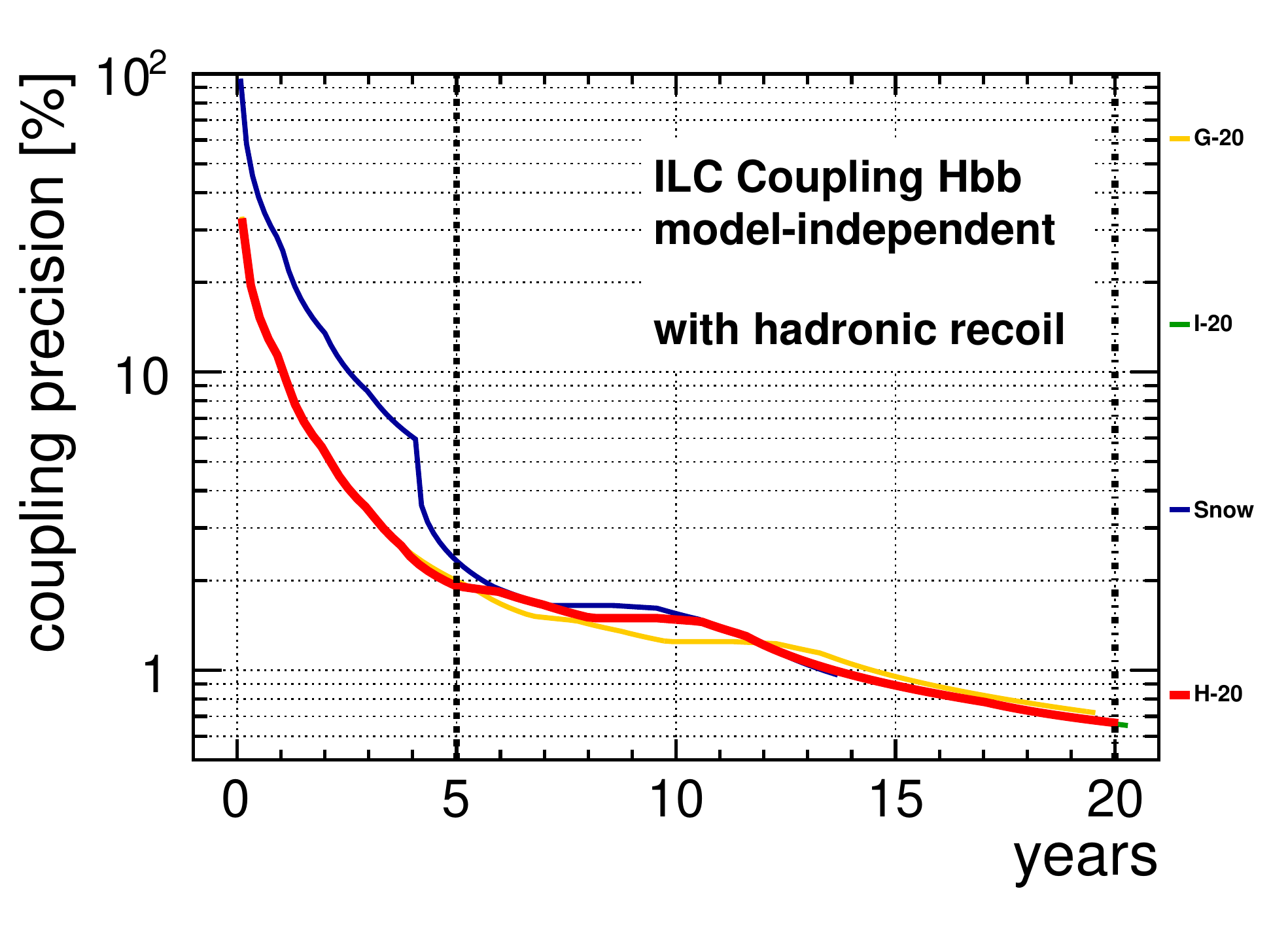}
\includegraphics[height=2.1in]{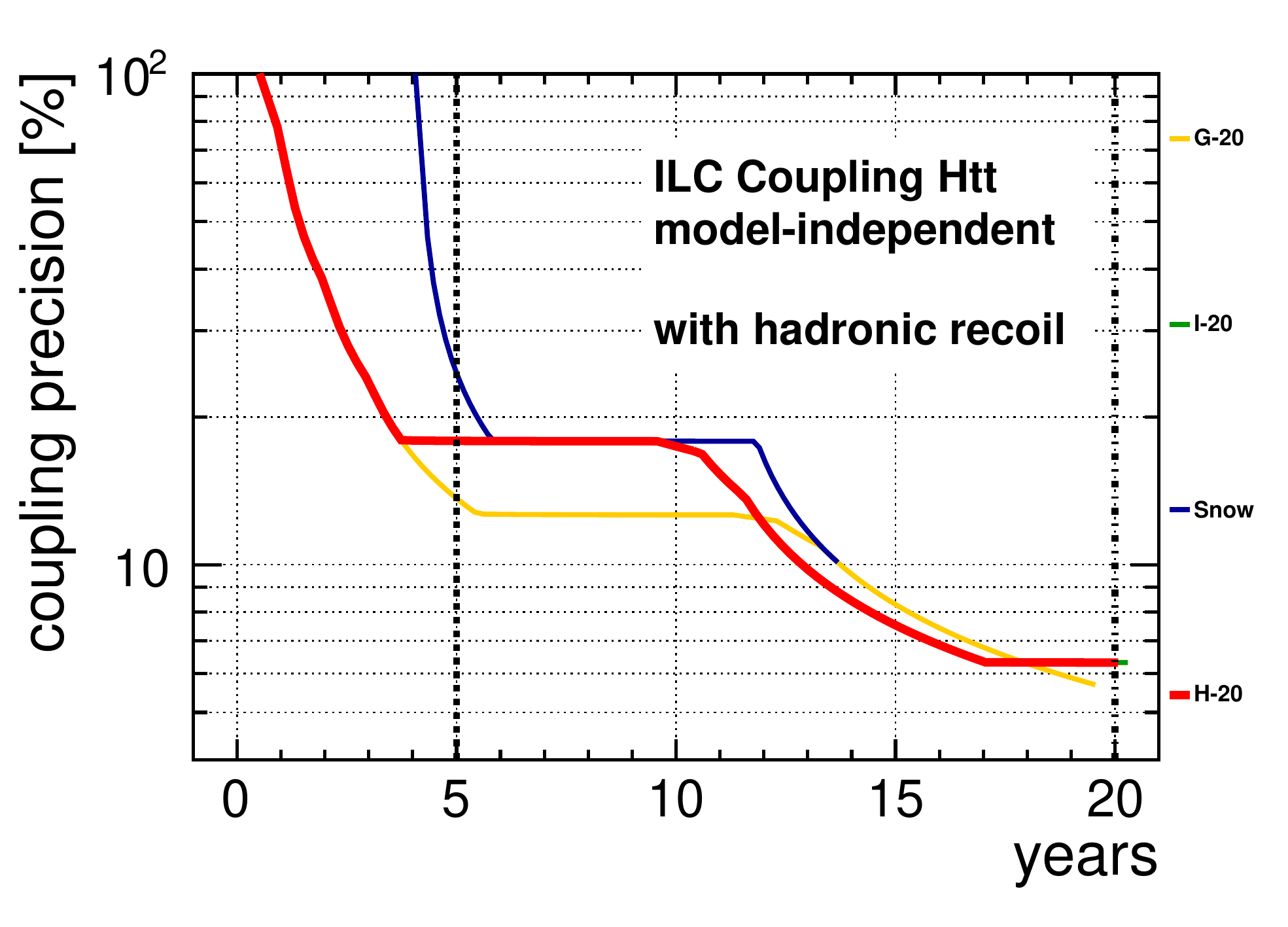}
\caption{Higgs coupling precision for the G-20, H-20, I-20, and Snowmass
scenarios.}
\label{fig:higgs}
\vspace{0.2 in}
\end{figure}

\begin{figure}[h!]
\begin{minipage}[h]{0.53\textwidth}
%\begin{figure}%[htb]
%\centering
%\includegraphics[height=4in]{higgs_WithHadrRec_GHISnow_logy_H-20_free}
\includegraphics[width=\textwidth]{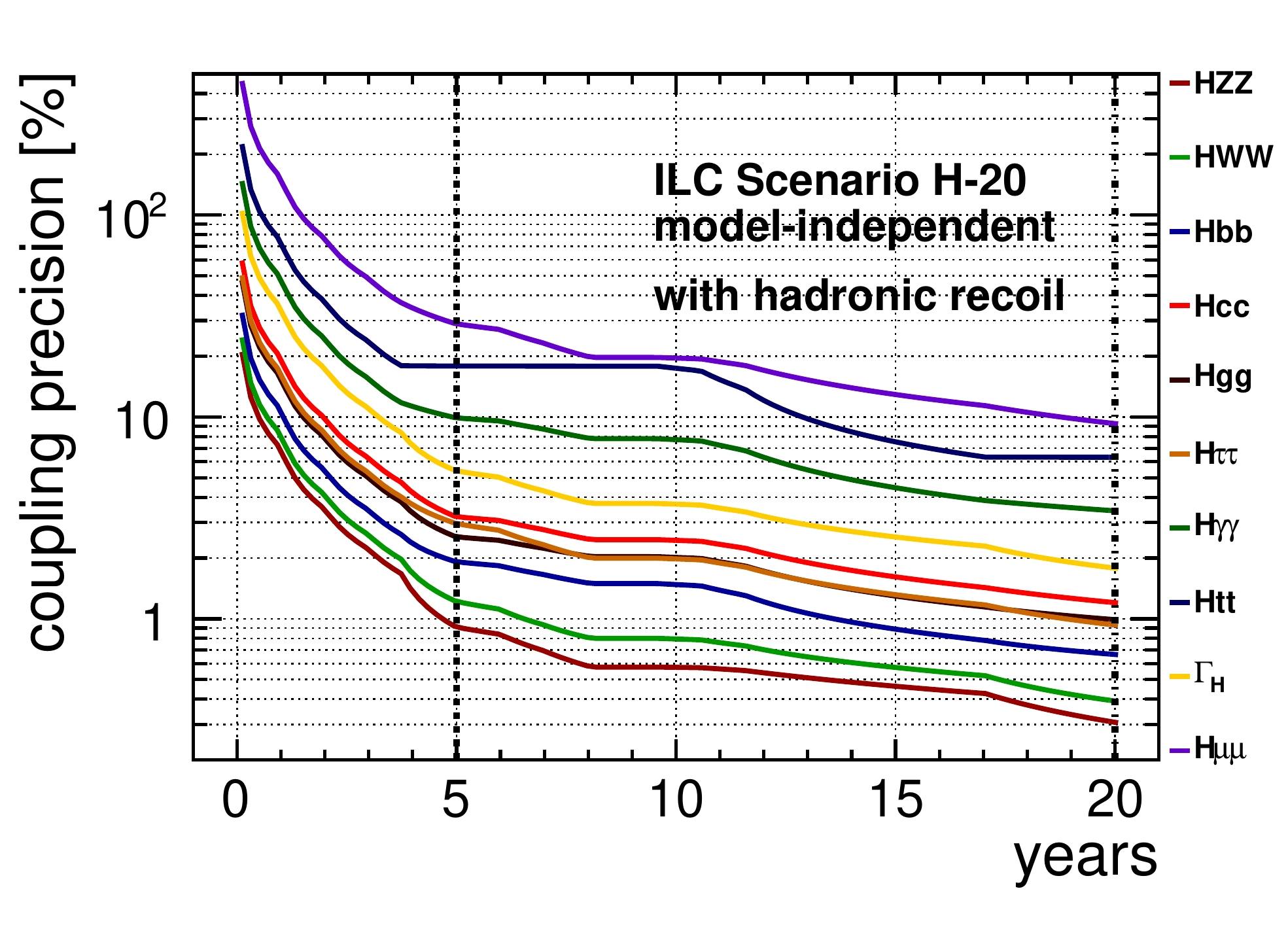}
\caption{Higgs coupling precision for H-20.}
\label{fig:higgs_h-20}
%\end{figure}
\end{minipage}
\hspace{.1 in}
\begin{minipage}[h]{0.47\textwidth}
%\begin{figure}[htb]
\centering
\includegraphics[width=\textwidth]{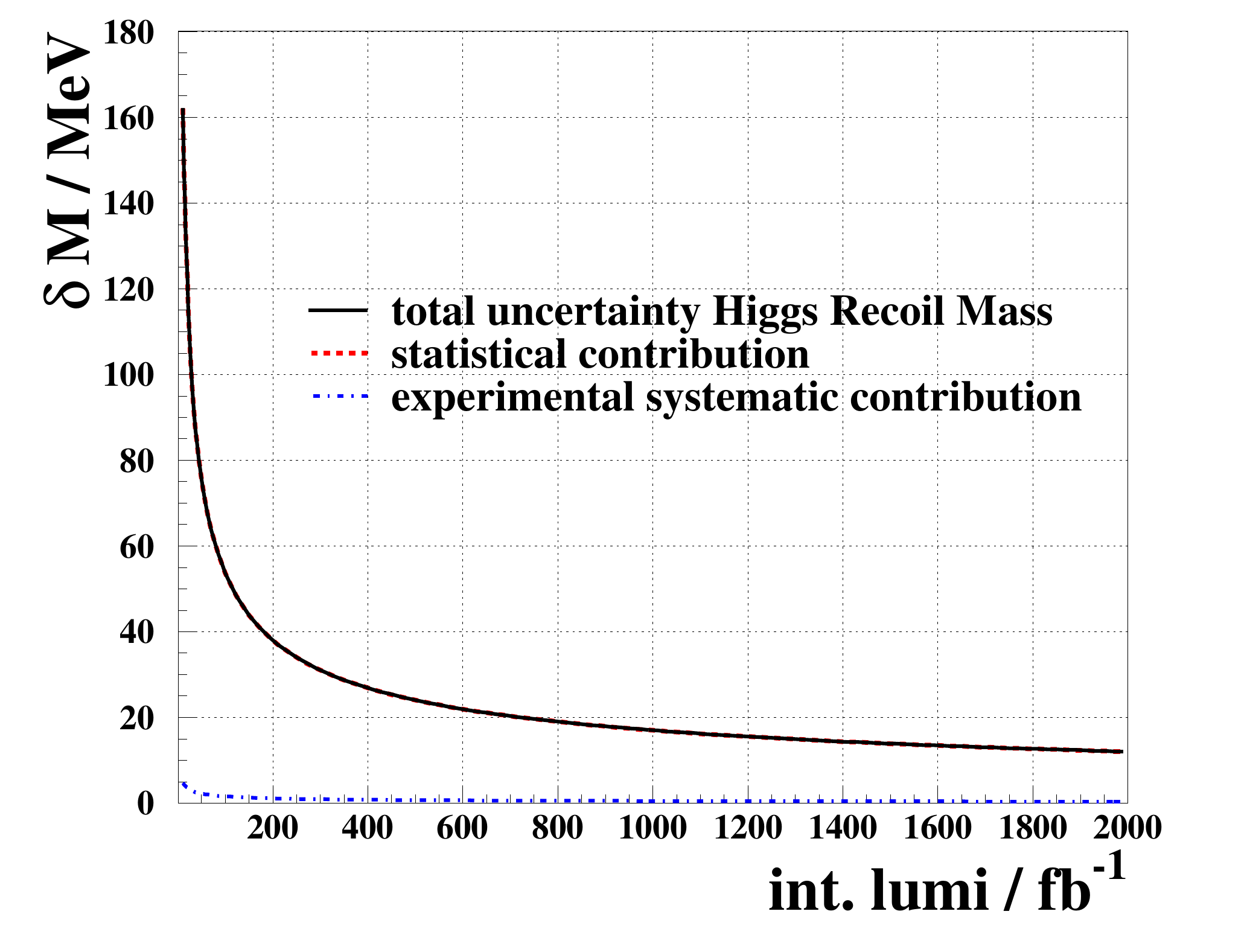}
\caption{Higgs mass precision versus integrated luminosity at $\sqrt{s}$ = 250 GeV.}
\label{fig:higgs_mass}
%\end{figure}
\end{minipage}
\end{figure}

Figure~\ref{fig:higgs} shows the resulting evolution of the Higgs coupling
precisions for HZZ, HWW, Hbb and Htt.
Comparison of the scenarios leads to the choice of H-20 
for its slightly better precision 
and longer 250 GeV operation, which may be needed
for the best Higgs mass measurement.
Figure~\ref{fig:higgs_h-20} shows the fuller set of couplings
measured in scenario H-20.
It must be emphasized that these precisions are model-independent.
The H-20 scenario has been approved by the Linear Collider Board (LCB)
as the official scenario to use in ILC physics projections~\cite{lcb-h20}.
Table~\ref{tab:lums} summarizes the total integrated luminosities
for this LCB-approved scenario.

%\section{Higgs boson mass}
The Higgs boson mass is a fundamental parameter of the Standard Model and
impacts the Higgs decay rates, for example $WW$ and $ZZ$,
through its couplings as well as the size of phase space.
Uncertainty on the Higgs mass leads to uncertainty in
the determination of couplings from measurements of
decay rates.  The LHC precision of about $\delta M_H = 200$\,MeV~\cite{Dawson:2013bba}
causes uncertainties of $2.2\%$ and $2.5\%$ on the partial widths of $H\to WW$ and $HH\to ZZ$, respectively~\cite{Asner:2013psa}, while an uncertainty of
$\delta M_H = 20$\,MeV is required to reach coupling uncertainties
 of $\sim 0.2\%$. 
Currently the only way demonstrated with full detector
simulation
to reach this level of precision 
 is the Higgs recoil mass measurement with $Z \to \mu \mu$ at $\sqrt{s}=250$\,GeV. With a momentum scale calibration from $Z\to\mu\mu$ at the $Z$ pole and an in-situ beam energy calibration from $\mu\mu\gamma$ events, systematic uncertainties should be controlled at the $1$\,MeV level~\cite{Graham}. Figure~\ref{fig:higgs_mass} shows the luminosity 
scaling of the Higgs recoil mass uncertainty. With $500$\,fb$^{-1}$ of data
collected at $\sqrt{s}=250$\,GeV, $\delta M_H = 25$\,MeV is reached.

\begin{table}%[t]
\begin{center}
\begin{tabular}{|c|c|c|c|}
\hline
  &first &after lumi&total  \\
  &phase &upgrade & \\ \hline
 250 GeV  &  500 fb$^{-1}$  &    1500 fb$^{-1}$   &   2 ab$^{-1}$ \\
 350 GeV  &  200 fb$^{-1}$  &      &   0.2 ab$^{-1}$  \\ 
 500 GeV  &  500 fb$^{-1}$  &  3500 fb$^{-1}$  & 4 ab$^{-1}$ \\ 
 \hline
 time   & 8.1 years  &  10.6 years &   20.2 years$^{*}$ \\  
 \hline
\end{tabular}
\caption{LCB-approved integrated luminosities for the ILC.
(*includes 1.5 years for luminosity upgrade.)}
\label{tab:lums}
\end{center}
\end{table}

\section{Top electroweak couplings}

The precision measurement of the electroweak couplings of the top quark is a key goal of the
ILC physics program. It requires beam polarization to disentangle the couplings to the $Z$ boson and the photon, which have different chiral properties. Besides being an important
test of the Standard Model, the top quark couplings are a prime indicator
for physics beyond the Standard Model. Due to the top quark's uniquely large mass, and thus its particularly strong coupling to the Higgs boson, new phenomena could
become visible first in the top sector.

\begin{figure}[htb]
%\centering
\begin{minipage}[t]{0.45\textwidth}
\includegraphics[width=3.1 in]{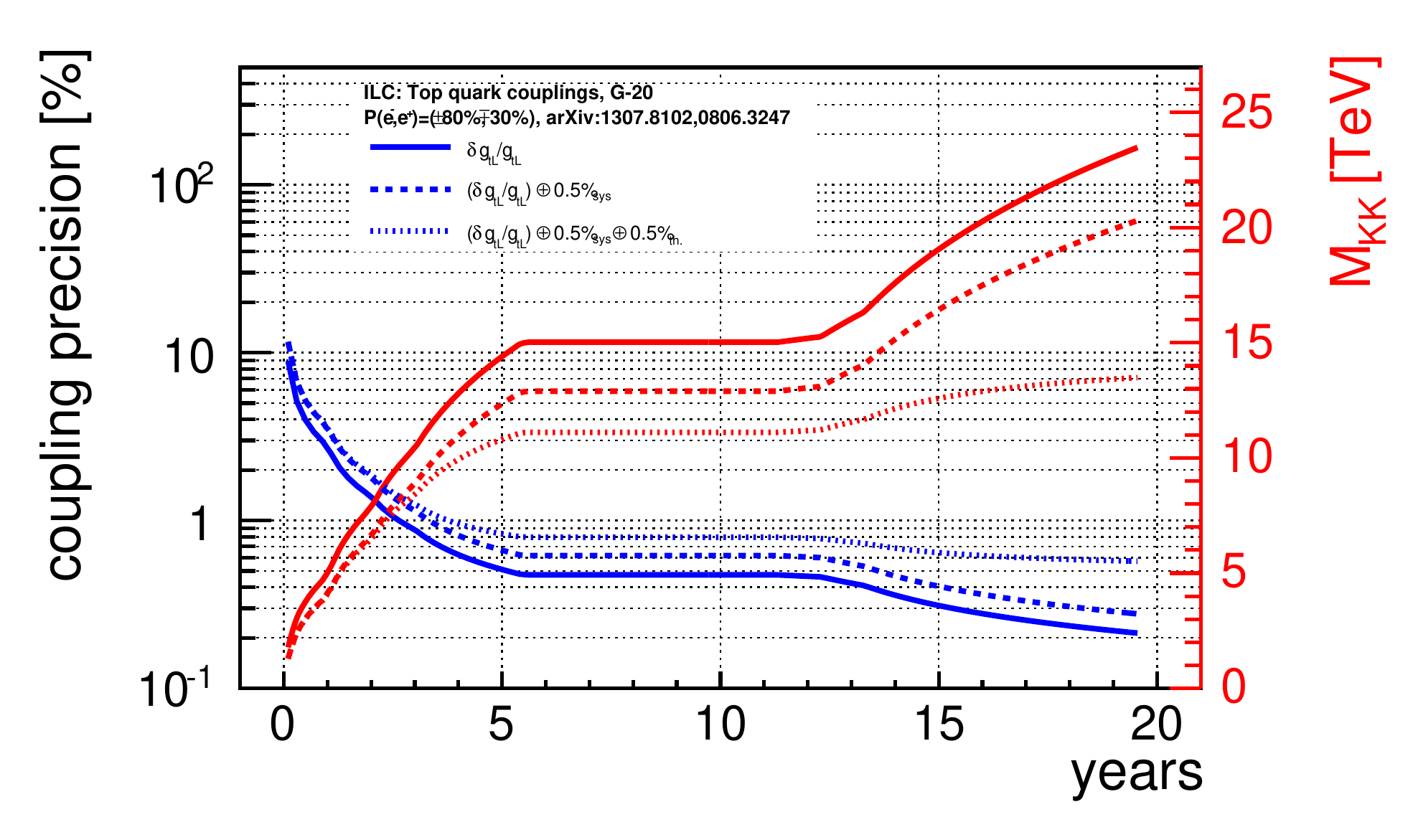}
\end{minipage}
\begin{minipage}[t]{0.2\textwidth}
\end{minipage}
\begin{minipage}[t]{0.45\textwidth}
\includegraphics[width=3.1 in]{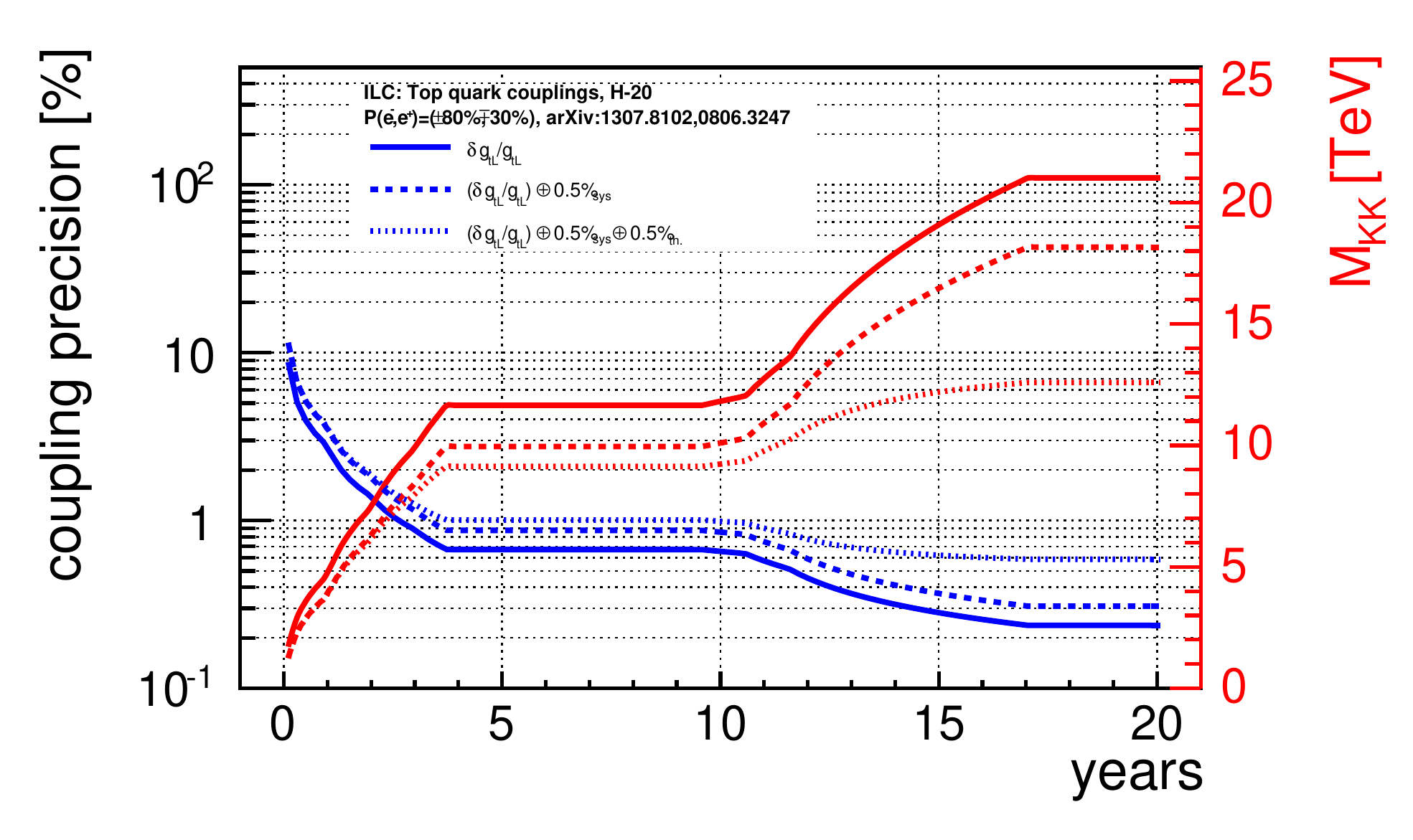}
\end{minipage}
\caption{Top electroweak left-handed couplings and the derived mass scale
sensitivity for Kaluza-Klein excitations in an extra-dimensions model
for scenarios G-20 (left) and H-20 (right).}
\label{fig:top_couplings}
\end{figure}

Figure~\ref{fig:top_couplings} shows the time evolution expected for 
the left-handed top coupling~\cite{Amjad:2013tlv}, and the sensitivity to the mass scale of new physics in an Extra-Dimension model derived by excluding deviations of the left-handed top coupling from its Standard Model prediction~\cite{Richard:2014upa}. In this model, indirect sensitivity for new physics can extend easily into the $10$-$15$\,TeV regime, which is beyond the reach of direct searches for resonances at the HL-LHC, estimated as $5$-$6$\,TeV~\cite{Agashe:2013hma}. The measurement of the electroweak couplings of the top quark
requires at least $\sqrt{s}>450$\,GeV.

\section{Higgs self-coupling}

An unambigous tree-level probe of the  Higgs self-coupling requires a measurement of the  double Higgs production cross section.  At the ILC, double Higgs production can be observed for $\sqrt{s}\ge 450$\,GeV;
this measurement is challenging and requires a large integrated luminosity.
A study based on full simulation of the ILD detector concept at $\sqrt{s}=500$\,GeV~\cite{Tian2013:LCNote}\cite{Duerig2014:privcom} using combined $HH \to b\bar{b}b\bar{b}$ and $HH \to b\bar{b}WW^*$ channels
has shown a precision of $30\%$ 
assuming an integrated luminosity of $4$\,ab$^{-1}$, shared equally between $P(e^-e^+)=(\pm 80\%,\mp 30\%)$.
Recently, improvements in the sensitivity of the
analyses have been identified.
Figure~\ref{fig:higgs_self} shows the time evolution of the precision on the
Higgs self-coupling for the scenarios G-20, H-20, I-20 and Snow. The helicities are chosen according to Table~\ref{tab:pols}. Before the luminosity upgrade, the precision is modest, but the full H-20 program reaches $27\%$~\cite{LCCPhysGroup:2015}. This would clearly demonstrate the existence of
the Higgs self-coupling.
%, particularly when combined with possible future LHC results~\cite{Dawson:2013bba}. 
The green line indicates the precision
that would be reached with the $1$\,TeV ILC upgrade,
where $10\%$ or better can be achieved.

The double Higgs production mechanisms at the two center-of-mass energies (500 GeV and 1 TeV)
 are different. The
sign of the interference term is different for double Higgsstrahlung and double Higgs production in $WW$-fusion. This means that a deviation of $\lambda$ from its Standard Model value will  lead to a larger
cross section for one process and a smaller cross section for the other. Thus the two measurements are complementary in their sensitivity to new physics. 

\section{Top Yukawa coupling}

The top Yukawa coupling is measured at the ILC
from the process $e^+e^- \rightarrow t \overline{t} h$, which
opens kinematically at around $\sqrt{s}=475$\,GeV.
Full detector simulation studies showed that at $\sqrt{s}=500$\,GeV,
the top Yukawa coupling can be determined with a precision of $9.9\%$ based
on an integrated luminosity of $1$\,ab$^{-1}$ with $P(e^-,e^+)=(-80\%,+30\%)$~\cite{Yonamine:2011jg}.
This translates
into final precision for H-20 of about $6\%$ (Figure~\ref{fig:higgs}).

\begin{figure}[t]
\centering
\begin{minipage}[t]{0.47\textwidth}
%\centering
\includegraphics[width=\textwidth]{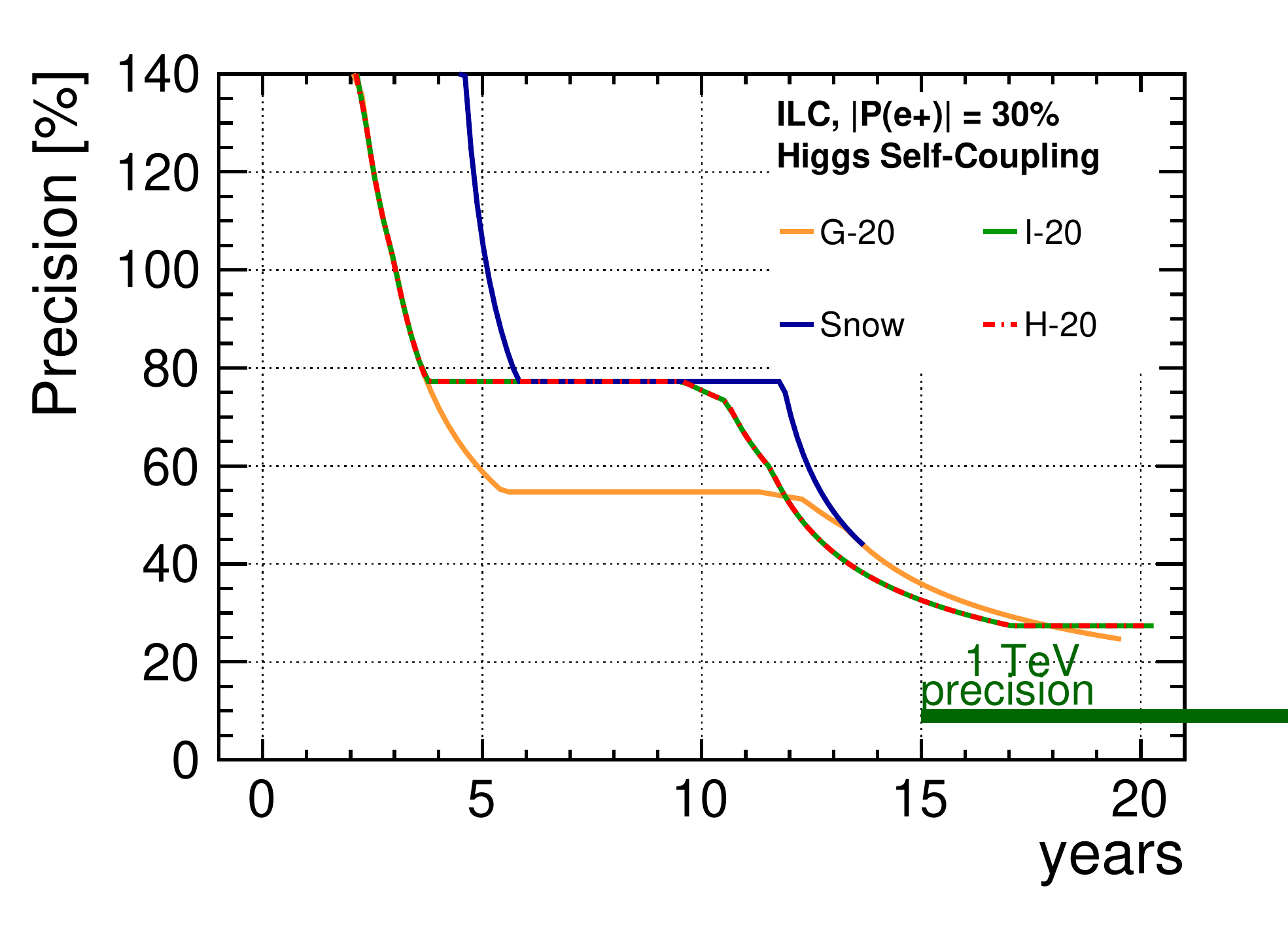}
\caption{Higgs self coupling. The ultimate precision for
the 1 TeV ILC is shown on the right.}
\label{fig:higgs_self}
%\end{figure}
\end{minipage}
\hspace{.15 in}
\begin{minipage}[t]{0.47\textwidth}
%\begin{figure}[htb]
%\centering
\includegraphics[width=\textwidth]{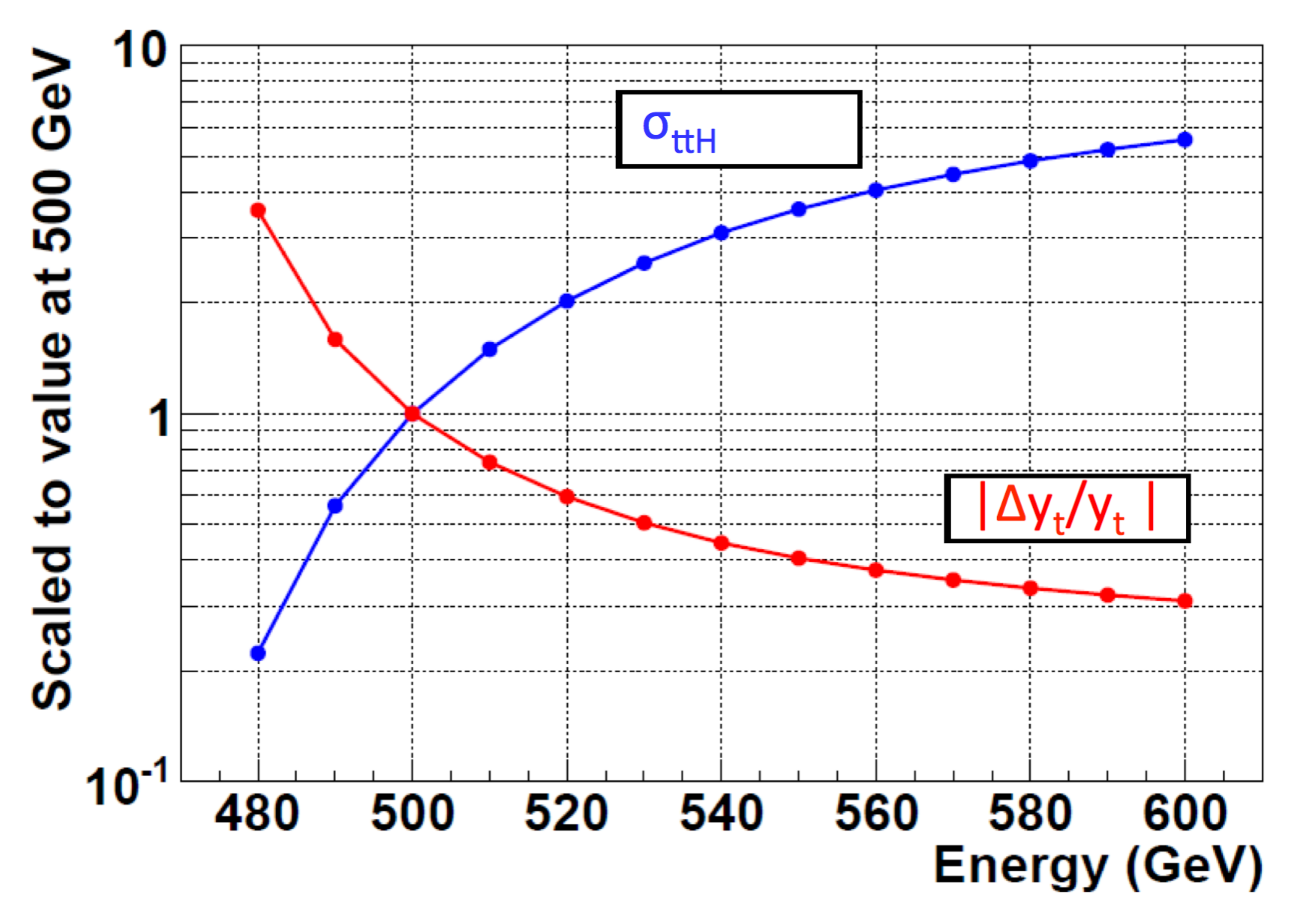}
\caption{tth cross section and top Yukawa coupling versus center-of-mass energy.}
\label{fig:top_yukawa}
\end{minipage}
\end{figure}

Figure \ref{fig:top_yukawa} presents the relative cross section for $t\bar{t}h$ production as
a function of $\sqrt{s}$; it is still steeply rising at $\sqrt{s}=500$\,GeV,
increasing nearly four-fold by $\sqrt{s}=550$\,GeV. Since the main
backgrounds (non-resonant $tbW$ and $t\bar{t}b\bar{b}$ production) decrease, the 
precision on the top Yukawa coupling improves by better than a factor of two w.r.t. 
$\sqrt{s}=500$\,GeV for the same integrated luminosity. This significant improvement in 
precision motivates serious consideration of extending the upper
center-of-mass reach of the nominally $500$\,GeV ILC to about $550$\,GeV.

It should also be noted that for $\sqrt{s}<500$\,GeV the
cross section drops quickly. For $\sqrt{s}=485$\,GeV, a reduction of $3\%$ in $\sqrt{s}$, 
the uncertainty would be twice as large as at $\sqrt{s}=500$\,GeV. 
Thus reaching at least $\sqrt{s}=500$\,GeV is essential to be
able to perform a meaningful measurement.

\section{Natural supersymmetry, light Higgsinos, and WIMP dark matter}

The motivations for physics beyond the Standard Model include the hierarchy problem
and dark matter. A possible solution to these mysteries is provided by natural supersymmetry,
including the possibility of light Higgsinos and WIMP dark matter candidates.  Should they exist,
the ILC offers valuable discovery potential. The highest available center-of-mass
energy as well as the possibility for threshold scans at lower energy are critical to
this potential.  The possibility to operate with all four helicity configurations
strengthens the role of the ILC in interpreting new particles.
Refer to the full report for details~\cite{scenarios}.

\section{Other operational details}
A number of additional operational issues have been considered. If new phenomena appear
at the LHC or the ILC the choice of running scenarios will be modified.  One strength
of the ILC is the ability to perform follow-up threshold scans for any such discovery.
Choices of beam helicity operations provide additional insight into the nature of
new physics.  The possibility of operating at WW-threshold or at the Z-pole may
prove to be important capabilities.  Each of these issues is discussed in~\cite{scenarios}. 

\section{Conclusions}
Based on studies of possible operating scenarios for the
500 GeV ILC a preferred scenario, H-20, has been identified.
Table~\ref{tab:lums} presents the assumed integrated luminosity for the 20-year program.
After starting operation at the full
center-of-mass energy of $500$\,GeV, 
running is planned at $250$ and $350$\,GeV before
the collider luminosity is upgraded for intense
running at $500$\,GeV and at $250$\,GeV.
Scenario H-20 optimizes the  possibility of discoveries of new physics while 
making the earliest measurements of the important Higgs properties. It includes
a sizeable amount of data taken at $\sqrt{s}=250$\,GeV.

The physics impact of the ILC is significantly improved if the
maximum energy of the $\sim 500$\,GeV ILC is stretched to $\sim 550$\,GeV where the
top Yukawa precision is more than a factor of two times better than at $500$\,GeV.

The choice of scenario H-20 is based on
the physics that is absolutely certain to be done with the ILC.
   This physics includes precision measurements of the Higgs boson and the top quark, and possibly measurements
   of the W and Z gauge bosons.  While this certain program provides a compelling and impactful scientific outcome,
   discoveries by the LHC or the early running of the ILC could expand the scientific impact
   of the ILC.  There exist scientific motivations to anticipate such possibilities.
   Such discoveries could alter the run plan from that described by H-20,
    as operations at or near the threshold of a pair-produced new particle, for example,
   would be added, a capability that is one of the particular operational strengths of the ILC.

\Acknowledgments
Many members of the ILC community contributed to this study through 
various studies and discussions, 
particularly Mikael Berggren, Roberto
Contino, Christophe Grojean, Benno List, Maxim Perelstein, Michael Peskin, Roman P\"oschl,
Juergen Reuter, Tomohiko Tanabe, Mark Thomson, Junping Tian, Graham Wilson and all members
of the LCC Physics Working Group.

\end{document}

%% file: econfmacros.tex
\def\beq{\begin{equation}}
\def\eeq#1{\label{#1}\end{equation}}
\def\eeqn{\end{equation}}

\def\beqa{\begin{eqnarray}}
\def\eeqa#1{\label{#1}\end{eqnarray}}
\def\eeqan{\end{eqnarray}}

\let\bar=\overbar

\def\Dslash{\not{\hbox{\kern-4pt $D$}}}
\def\dslash{\not{\hbox{\kern-2pt $\del$}}}

\def\msb{{\bar{\ssstyle M \kern -1pt S}}}